\documentclass[twocolumn,english,prl,twocolumn]{revtex4-1}
\usepackage{lmodern}
\usepackage{lmodern}
\usepackage[T1]{fontenc}
\usepackage[latin9]{inputenc}
\usepackage{graphicx}
\usepackage{esint}
\usepackage{upgreek}
\makeatletter

\usepackage{verbatim}
\usepackage{amstext}
\usepackage{babel}

\makeatother

\usepackage{babel}
\def\mr{\mathrm}
\def\mi{\mathit}

\begin{document}

\title{Mis-orientation controlled cross-plane thermoelectricity in twisted bilayer graphene}

\author{Phanibhusan S. Mahapatra$^{1,\ast}$, Bhaskar Ghawri$^{1,\#}$, K. Watanabe$^{3}$, T. Taniguchi$^{3}$, Subroto Mukerjee$^{1}$  \& Arindam Ghosh$^{1,2}$}

\affiliation{$^{1}$Department of Physics, Indian Institute of Science, Bangalore
560 012, India. }
\affiliation{$^{2}$Centre for Nano Science $\&$ Engineering, Indian Institute of Science, Bangalore
560 012, India. }
\affiliation{$^{3}$National Institute for Materials Science, Namiki 1-1, Tsukuba, Ibaraki 305-0044, Japan. }

\begin{abstract}
Introduction of \lq twist' or relative rotation between two atomically thin van der Waals (vdW) membranes gives rise to periodic Moir\'{e} potential, leading to a substantial altercation of the band structure of the planar assembly. While most of the recent experiments primarily focus on the electronic-band hybridization by probing in-plane transport properties, here we report out-of-plane thermoelectric measurements across the van der Waals gap in twisted bilayer graphene (tBLG), which exhibits an interplay of twist-dependent inter-layer electronic and phononic hybridization. We show that at large twist angle, the thermopower is entirely driven by a novel phonon drag effect at sub-nanometer scale, while the electronic component of the thermopower is recovered only when the mis-orientation between the layers is reduced to $\lesssim 2^{\circ}$. Our experiment shows that cross-plane thermoelectricity at low angle is exceptionally sensitive to nature of band dispersion and may provide fundamental insights to the coherence of electronic states in twisted bilayer graphene.

\end{abstract}

\maketitle
The van der Waals (vdW) interaction between two graphene membranes in a Moir\'{e} super-lattice (Fig.~1a) can be precisely manipulated by a relative rotation or twist between the constituent layers \cite{koren2016coherent,cao2018correlated,cao2018unconventional,mele2010commensuration,perebeinos2012phonon,kim2013breakdown,dos2007graphene,bistritzer2010transport,bistritzer2011moire,yankowitz2019tuning}. At small twist angle, $\mathrm{\theta < 3^{\circ}}$, the strong inter-layer hybridization can alter the low-energy super-lattice band structure significantly, leading to a phase-coherent tunneling of electrons across the layers with a renormalized Fermi velocity \cite{fang2016electronic,luican2011single,cao2016superlattice,EnergySpectrum2012Moon,bistritzer2011moire}. The phase coherence of the inter-layer tunnelling is maintained as long as the tunnelling time scale ($\hbar/\gamma$, where $\gamma$ is the inter-layer coupling) is smaller than the in-plane dephasing time scale ($\tau$)  \cite{bistritzer2010transport}. In contrast, the inter-layer hybridization for large twist angle, ${\theta > 3^{\circ}}$, occurs at higher energies and hence at higher doping, leaving the low energy bands of the two layers essentially decoupled at low temperature ($T$). As a result, electrons tunnel incoherently across the vdW gap as the two successive tunnelling events no longer remain phase-coherent ($\hbar/\gamma \gg \tau$)\cite{kim2013breakdown,bistritzer2010transport,luican2011single}. However, at higher temperature regime set by the Bloch-Gru$\ddot u$neisen temperature, $T > T_{\mr{BG}}= 2\hbar v_\mr{ph} k_\mr{F}/k_\mr{B} $, where $k_\mr{B}$,~$v_\mr{ph}$ and $k_\mr{F}$ are the Boltzmann constant, phonon velocity and the Fermi wave vector, respectively, the low energy quasiparticle excitations are coupled with the layer breathing modes of phonons (LBM) through electron-phonon (e-ph) scattering \cite{mahapatra2017seebeck,perebeinos2012phonon}. The phonon-mediated recoupling of the two layers  manifests in an unconventional phonon-drag effect in the thermoelectric transport between two atomically thin layers \cite{mahapatra2017seebeck}. This completely phonon-driven thermoelectric transport persists even at low temperature, suggesting that the electric and thermoelectric transport coefficients cannot be related by the semiclassical Mott relation for conventional tunnel junctions \cite{mahapatra2017seebeck,sadeghi2016cross,hung2014enhanced}. However, the relevance of layer-hybridized phonons in the thermoelectric transport remain unclear when the electronic hybridization of the two layers becomes strong at low $\theta$. And a systematic experimental study on twist angle dependence is needed to develop the physics of the inter-layer energy transport when the van der Waals interface is subjected to a statistical driving force by establishing a temperature gradient.

\begin{figure}[t]
\includegraphics[clip,width=9cm]{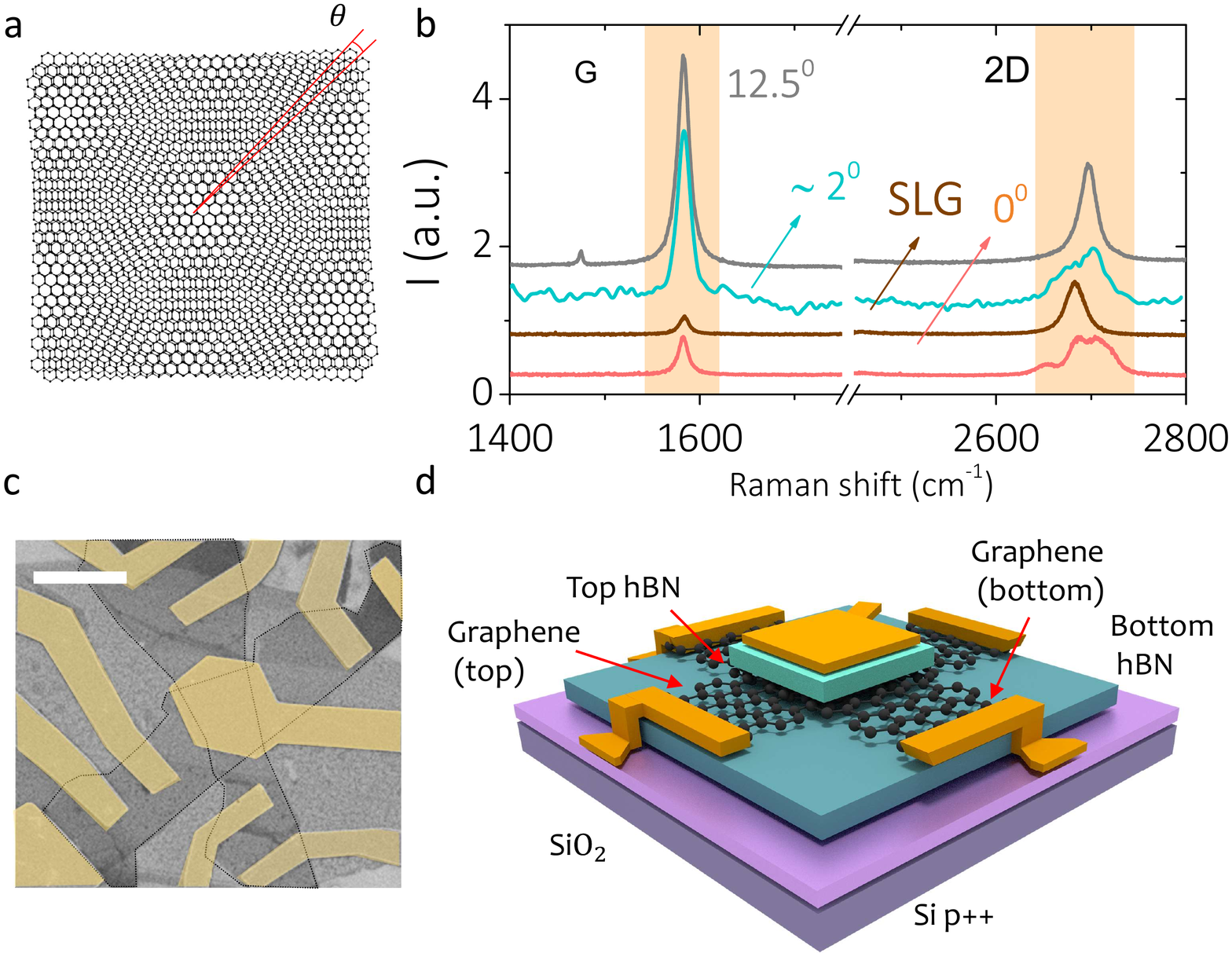}\caption{\textbf{ Device structure and characterization}: (a) Moir\'{e} super-lattice when a relative rotation ($\theta$) is introduced between two graphene layers. (b) The Raman spectra for $G$ peak and $2$D peak (shaded region) are compared for $\theta \sim 12.5^{\circ}$, $\theta \sim 2^{\circ}$, $\theta = 0^{\circ}$ and single layer graphene with relative offset in the intensity for clarity. (c) Scanning electron microscopy (SEM) image of a device with twist angle $\theta = 0^{\circ}$ (Bernal stacking). The scale bar represents a length of $5$~$\mu$m. (d) Device schematic for the cross-plane electrical and thermoelectric measurements. }
\end{figure}

In this letter, we report the measurement of thermoelectric transport across a single vdW gap formed in twisted bilayer graphene. To have independent access to both the layers as well as the cross-junction, we create the van der Waals stack of two graphene layers at $60^{\circ} + \theta$, where $\theta$ is the specific mis-orientation angle. The graphene super-lattice is then encapsulated within two hexagonal Boron Nitride (hBN) layers in a vertical stack on Si/SiO$_2$ substrate (see Fig.~1d). We have measured a total of six devices of which four have large $\theta > 3^{\circ}$, one device with Bernal/AB stacking ($\theta = 0^{\circ}$) and one device with $\theta \sim 2^{\circ}$. The observed difference in the Raman spectra from the mono-layer graphene and overlap region suggests the twist angle, $\theta \sim 6^{\circ}, 10^{\circ}, 12.5^{\circ}$ and $14^{\circ}$ for the devices with large $\theta$ (see supplementary information for more details). The device with  $\theta \sim 2^{\circ}$ shows a much broader shape of the $2$D peak with additional shoulder while the Bernal stacked device exhibits a $2$D peak with four-component characteristic structure of $2$D band (Fig.~1b) similar to exfoliated bilayer graphene flakes \cite{ferrari2006raman,malard2009raman}. The doped Si/SiO$_2$ substrate acts as a global back gate while a local top gate on the overlap region controls the doping density of the overlap region independently as shown in the scanning electron microscopy (SEM) image in Fig.~1c. 
\begin{figure}[t]
\includegraphics[clip,width=9cm]{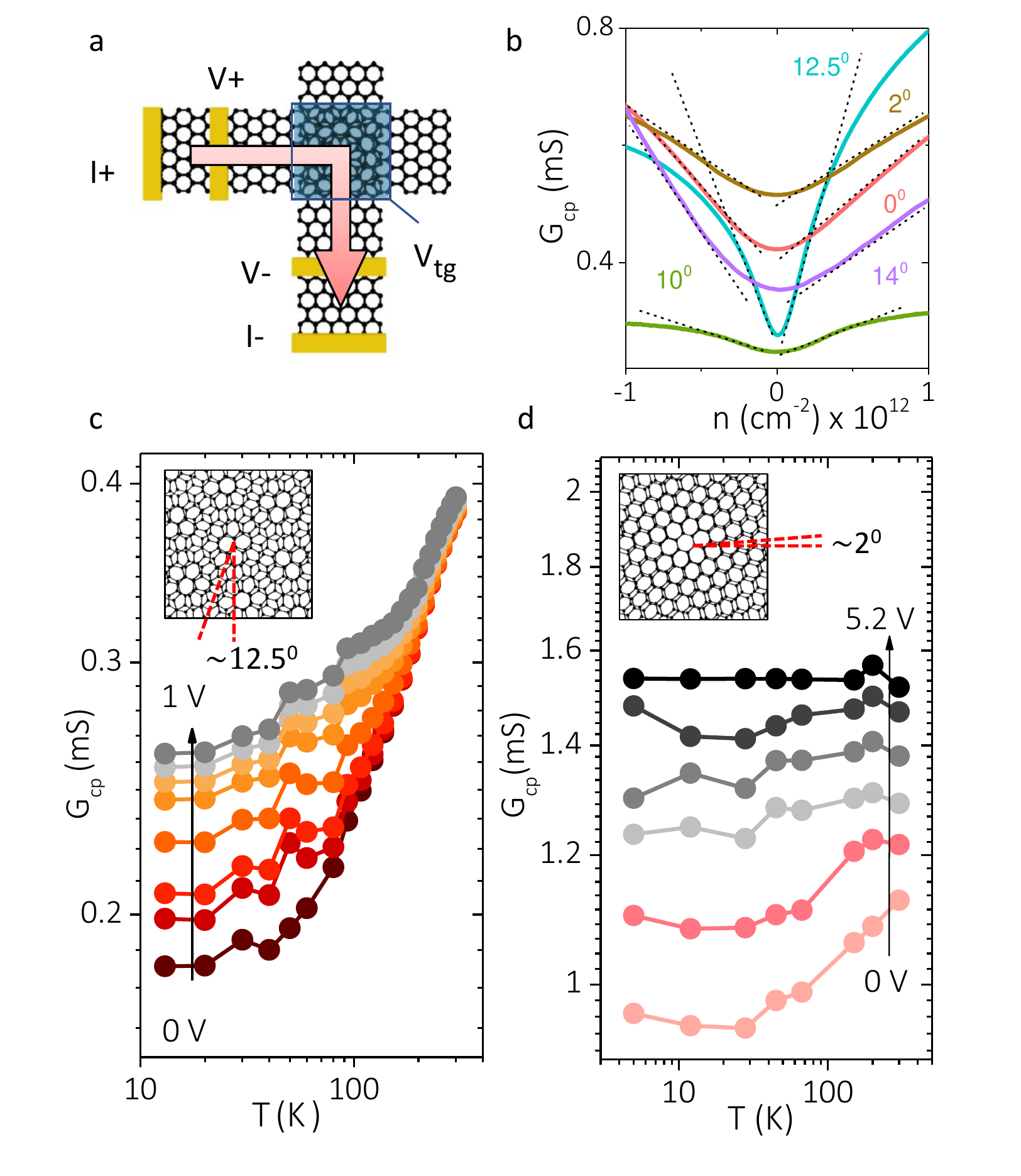}\caption{\textbf{Cross-plane electrical transport}: (a) Schematic for the four-terminal cross-plane conductance measurements. (b) The density dependence of cross-plane conductance $G_{\mr{cp}}$ is shown for various twist angles ($\theta$) at room temperature with relative offset in the magnitude of $G_\mr{cp}$ for clarity. The dotted lines denote the linearity of $G$ with number density ($n$) at low $n$. Temperature dependence of $G_{\mr{cp}}$ for (c) $\theta \sim 12.5^{\circ}$ and (d) $\theta \sim 2^{\circ}$, respectively, for different values of doping ($|V_\mr{tg}-V_\mr{D}|$).}
\end{figure}

Fig.~2a depicts the measurement schematic for the four-terminal cross-plane conductance ($G_\mr{cp}$). The local top gate potential $|V_\mr{tg}-V_\mr{D}|$, where $V_\mr{D}$ is the Dirac point position, controls the doping in the tBLG region while the back gate is maintained at a high potential ($V_{\mr{bg}} \approx -35$~V) to minimize the series contribution from the monolayer region in the current path. The cross-plane charge transport can be driven by two distinct processes: ($1$) inter-layer charge tunnelling and ($2$) phonon-assisted charge transfer \cite{mahapatra2017seebeck,kim2013breakdown,perebeinos2012phonon}. Both processes exhibit Drude conductivity relation $G_{\mr{cp}} \propto n$, where $n$ is the carrier density, at low $n$ (Fig.~2b). For large $\theta$, the inter-layer conduction at low temperature ($T \lesssim 70$~K) originates from incoherent tunnelling between the two graphene layers, leading to $T$-independent $G_{\mr{cp}}$ \cite{kim2013breakdown,bistritzer2010transport}. However, at higher temperature, the LBM phonons assist in inter-layer conduction through e-ph scattering, leading to an increasing $G_{\mr{cp}}$ with temperature as shown in Fig.~2c for $\theta \sim 12.5^{\circ}$ \cite{perebeinos2012phonon,kim2013breakdown,mahapatra2017seebeck}. The cross-over temperature, $T' \sim 70$~K is characterised by the onset of phonon-driven transport and provides an energy scale of the interlayer phonons.

\begin{figure}[t]
\includegraphics[clip,width=9cm]{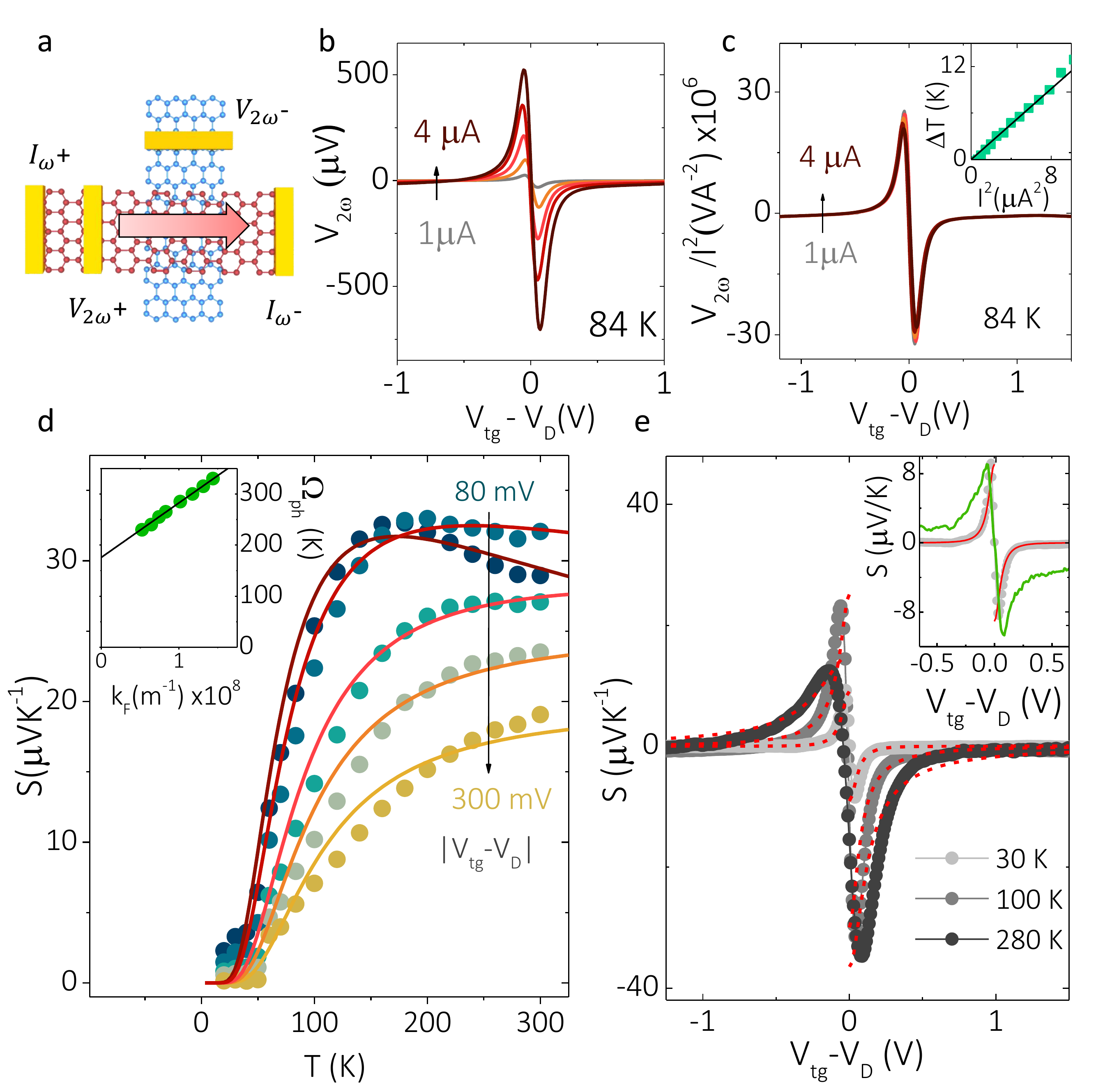}\caption{\textbf{Thermoelectric transport at large twist angle ($\theta \sim 12.5^{\circ}$ )}: (a) In-plane heating and the measurement scheme for cross-plane thermovoltage $V_{2\omega}$ . (b) $V_{2\omega}$ with varying top-gate voltages $|V_\mr{tg}-V_\mr{D}|$ for different in-plane heating currents ($1-4$~$\mu$A) at $84$~K. (c) $V_{2\omega}$  normalized with $I_{\omega}^2$. The inset shows that the measured temperature difference $\Delta T \propto I_{\omega}^2$. (d) Temperature dependence of $S = V_{2\omega}/\Delta T$ for $\theta \sim 12.5^{\circ}$ device for various top gate voltages $|V_\mr{tg}-V_\mr{D}|$. The solid lines show the fit of the phonon-driven thermopower (TEP) described in Eq.~\ref{phonondrag}. The inset shows the obtained phonon energy as a function of Fermi wave-vector $k_\mr{F}$. (e) The density dependence of the measured $S$ for three representative temperatures (circles). The dashed lines show the fitted $S$ from the phonon drag TEP (Eq.~\ref{phonondrag}). The inset shows the comparison between the measured $S$ (grey line) and the calculated $S$ (green line) from the Mott relation (Eq.~\ref{Mott relation}) at the lowest temperature $T=30$~K. The red lines show the fit of the phonon-drag mediated TEP. }
\end{figure}

The $T$-dependence of cross-plane conductance in both $\theta \sim 2^{\circ}$ and Bernal stacking is distinctly different from that at large $\theta$, since $G_{\mr{cp}}$ is almost temperature independent as shown for $\theta \sim 2^{\circ}$ in Fig.~2d, suggesting the absence of phonon-mediated scattering in cross-plane conduction. We also note that while the recent experiments show large linear-in-$T$ in-plane resistivity at low twist angle ($\theta \leq 2^{\circ}$) due to strongly enhanced electron-phonon scattering \cite{Polshyn2019large}, such quasi-elastic phonon scattering processes are most likely filtered out in the cross-plane transport due to momentum conservation.

The cross-plane thermoelectric power (TEP) or Seebeck coefficient is obtained from $S(V_\mr{tg},T)= V_{2\omega}/\Delta T$, where $\Delta T$ is the effective temperature difference between the two graphene layers, created by passing a sinusoidal heating current ($I_{\omega}$) in the top graphene layer (Fig.~3a). The resulting second harmonic thermo-voltage $V_{2\omega}$ is recorded between the two layers for various doping (Fig.~3b) \cite{mahapatra2017seebeck,zuev2009thermoelectric}.  At a particular $T$, the maximum of $|V_{2\omega}|$ reaches at a $V_\mr{tg}$ value symmetrically close to the CNP and sharply decreases with increasing doping, a behavior qualitatively similar to the in-plane graphene TEP \cite{zuev2009thermoelectric,checkelsky2009thermopower}. The cross-plane $\Delta T$ is measured using resistance thermometry of the top graphene layer (See supplementary information for more details).  For the range of heating current used,  $I_{\omega} \sim 1-4$~$\mu$A, both $V_{2\omega}$, $\Delta T \propto I_{\omega}^2$ (Fig.~3c) ensure that the measurements were performed within the linear response regime ($\Delta T \ll T$) and the Seebeck coefficient $S$ is independent of the $\Delta T$ itself. 

First, we compare the temperature dependence of $S$  at various doping (\textit{i.e.,} $|V_\mr{tg}-V_\mr{D}|$) for $\theta \sim 12.5^{\circ}$ in Fig.~3d. The measured $S$ exhibits a nonlinear $T$-dependence as observed in our previous work \cite{mahapatra2017seebeck} which suggests that the cross-plane TEP is primarily driven by the inter-layer phonons. A phonon-driven TEP involves temperature difference ($\Delta T$)-induced quasi-non-equilibrium condition which leads to net diffusion of phonons from the hot layer to the cold layer. These out-of-equilibrium phonons then impart momentum to the charge carriers through e-ph scattering, leading to a frictional drag force \cite{wu1996phonon,cantrell1987calculation} on the charge carriers.
In the steady state, this phonon drag force results in additional thermal voltage between the two layers due to the inter-layer charge imbalance. For the quadratic dispersion relation of LBM branch of phonons \cite{perebeinos2012phonon, mahapatra2017seebeck}, the phonon drag component of TEP can be  expressed as \cite{wu1996phonon}   

\begin{equation}
S \approx \frac{\alpha \mi{\Omega}_{\mr{ph}}}{neT^{2}}\frac{e^{\mi{\Omega}_{\mr{ph}}/T}}{(e^{\mi{\Omega}_{\mr{ph}}/T}-1)^{2}}
\label{phonondrag}
\end{equation}

where $n$ is the number density of the carriers with charge $e$ and the pre-factor $\alpha$ captures different phonon scattering rates. Here $\mi{\Omega}_{\mr{ph}}(\mathbf{q}_{K},k_{\mr{F}})$ is the energy of the LBM phonon that elastically scatters one electron from Fermi circle of ($k_\mr{F}$) one layer to another which is separated by momentum $\mathbf{q}_{K}$ (for more details see supplementary information).

The phonon-drag mediated TEP in Eq.~\ref{phonondrag} shows excellent fit to the $T$-dependence of $S$ at various doping density for $\theta \sim 12.5^{\circ}$ as shown in Fig.~3d. The fitting parameter $\alpha$ is found to be temperature independent for higher doping but becomes weakly temperature dependent $\sim T^{-\gamma}$, where $\gamma \approx 0.1 - 0.3$, close to CNP . The fit of Eq.~\ref{phonondrag} to the $T$-dependence of TEP yields $\mi{\Omega}_{\mr{ph}} (\mathbf{q}_{K},k_{\mr{F}}) \sim 200 - 300$~K and exhibits linear dependence on Fermi wave vector $k_{\mr{F}}$ (inset Fig.~3d). This is the direct consequence of momentum conservation in the inter-layer e-ph scattering as the average phonon momentum required to scatter one electron from one layer to another $\approx \mathbf{q}_{K}+k_{\mr{F}}$ \cite{mahapatra2017seebeck}. The intercept $\mi{\Omega}_{\mr{ph}} (\mathbf{q}_{K},k_{\mr{F}}=0) \approx 175$~K coincides well with the low-energy LBM branch ZO$^{'}$/ZA$_{2}$ in tBLG \cite{cocemasov2013phonons,campos2013raman,mahapatra2017seebeck}.

The density dependence of $S$, shown in Fig.~3e (red dotted lines) for three different temperatures, can also be obtained quantitatively from electron-phonon scattering scenario. In fitting the phonon drag TEP, we have used Eq.~\ref{phonondrag} and the linear dependence of $\mi{\Omega}_{\mr{ph}}$ on $k_{\mr{F}} (= \sqrt{\pi n})$ with a charge-puddle broadening factor ($n_0$) in the density $n$ such that $\mi{\Omega}_{\mr{ph}}= \Omega_0 + \beta \sqrt{\pi \big(n_0^2 + n^2 \big)^{\frac{1}{2}}}$, where the coefficients $\Omega_0$ and $\beta$ are taken from the linearity of $\mi{\Omega}_{\mr{ph}}$ on $k_\mr{F}$.

In contrast to the large $\theta$ devices, TEP in both $\theta \sim 2^{\circ}$ and $\theta = 0^{\circ}$ devices (Fig.~4a \& 4b) exhibits a linear dependence on temperature throughout the experimental temperature range ($\sim 30-300$~K). This is qualitatively similar to the  Mott-thermopower ($\sim T/T_\mr{F}$) observed for graphene in-plane and diffusive conductors in the degenerate limit $k_\mr{B}T \ll \mu$, where $\mu$ is the chemical potential \cite{zuev2009thermoelectric,theoryofthermopower}. While the origin of TEP $\propto T$ can be attributed to the inter-layer entropy transport by the thermally activated quasi-particles over the Fermi energy, the absence of nonlinearity in $T$-dependence of $S$ also indicates that the phonon drag effects are negligible in low twist angles irrespective of $T$. Interestingly, for $\theta = 0^{\circ}$, we observe the linear scaling of $S$ with respect to $T$ even at small carrier densities $\sim 6 \times 10^{11}$~cm$^{-2}$, where the degenerate limit is not expected to hold due to the finite DOS of AB stacked bilayer graphene close to CNP \cite{nam2010thermoelectric}. However, the non-degenerate regime is difficult to realize in a non-ideal device where spatial charge inhomogeneities rather than the gate potential set the minimum carrier density ($n_0$) close to CNP, which we estimate $\sim 4-6 \times 10^{11}$~cm$^{-2}$ at $300$~K (see supplementary information). We also note that the absolute magnitude of $|S|$ for $\theta = 0^{\circ}$ only reaches $\sim 1-5$~$\mu$VK$^{-1}$ which is one order less compared to large $\theta$ tBLG devices \cite{mahapatra2017seebeck}. The observed suppression of $S$ is presumably due to the branching of heating current into the other graphene, as the two layers are strongly hybridised, leading to an overestimation of measured $\Delta T$ from the resistance thermometry.

\begin{figure}[t]
\includegraphics[clip,width=9cm]{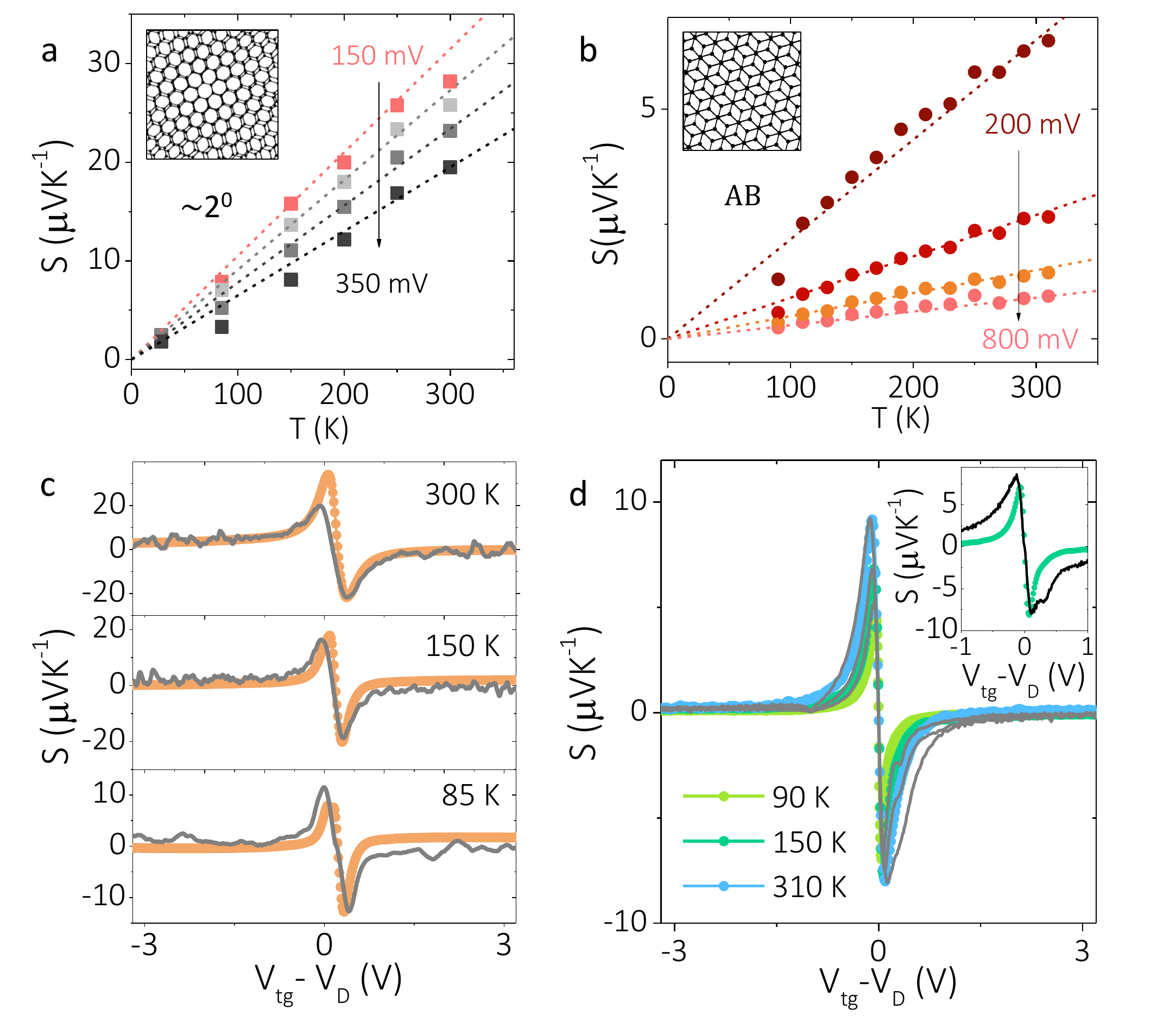}\caption{\textbf{Cross-plane thermoelectricity at low twist angle}: Temperature dependence of $S$ for (a) $\theta \sim 2^{\circ}$ and (b) $\theta = 0^{\circ}$ (Bernal stacking) for various top gate voltages $|V_\mr{tg}-V_\mr{D}|$. The dashed lines show the linear $T$-behaviour as a guide for the eye. The doping dependence of experimentally measured $S$ (coloured circles) is compared with $S_{\mr{Mott}}$ (grey solid lines) for (c) $\theta \sim 2^{\circ}$ and (d) $\theta = 0^{\circ}$ at various temperatures. $S_{\mr{Mott}}$ is calculated using the (c) single layer Dirac dispersion and (d) parabolic band dispersion of bilayer graphene. The inset in (d) shows the density dependence of $S$ for $\theta = 0^{\circ}$ compared to that of normalized $S_{\mr{Mott}}$~(black line) evaluated from the single layer Dirac dispersion at a fixed representative temperature $T=150$~K. }
\end{figure}

We now analyse the doping dependence of $S$ in both $\theta \sim 2^{\circ}$ and $\theta = 0^{\circ}$ devices considering that the TEP is composed of the electronic component and hence determined by the Mott relation \cite{jonson1980mott,zuev2009thermoelectric},
\begin{equation}
S_{\mr{Mott}}=-\frac{\pi^{2}k_{\mr{B}}^{2}T}{3|e|}\frac{1}{G_\mr{cp}}\frac{dG_{\mr{cp}}}{dV_\mr{tg}} \frac{dV_\mr{tg}}{dE}\bigg\vert_{E=E_{\mr{F}}}
\label{Mott relation}
\end{equation}
 $S_{\mr{Mott}}$ in Eq.~\ref{Mott relation} can be evaluated by differentiating the experimentally measured $G_{\mr{cp}}$ with respect to $V_\mr{tg}$ and using the parallel plate model of gate capacitance. For $\theta \sim 2^{\circ}$, the Dirac dispersion of single layer graphene (SLG), $E_\mr{F} = \hbar v_\mr{F} \sqrt{\pi n}$, yields  $\frac{dV_{\mr{tg}}}{dE} = \frac{2}{\hbar v_\mr{F}} \sqrt{\frac{2e}{\pi C_\mr{BN}}|V_\mr{tg}-V_\mr{D}|} $ , where $v_\mr{F}= 10^6$~ms$^{-1}$ and $C_\mr{BN}$ are the Fermi velocity in SLG and gate capacitance per unit area, respectively. In calculating $\frac{dV_{\mr{tg}}}{dE}$, we have assumed that the gate potential $|V_\mr{tg}-V_\mr{D}|$ induces equal doping density in both layers due to negligible screening of electric field from the graphene sheet and small inter-layer separation $d \sim 0.4$~nm \cite{kim2013breakdown,sanchez2012quantum}. We compare the doping dependence of TEP in $\theta \sim 2^{\circ}$ with Mott Relation at various temperatures in Fig.~4c. $S_\mr{Mott}$ obtained from Eq.~\ref{Mott relation} coincides well with the measured $S$ for the three representative temperatures. The quantitative agreement of $S$ with $S_\mr{Mott}$ also suggests that the re-normalization effects on $v_{F}$ due to the flattening of lowest-energy bands are not significant in our device \cite{luican2011single}.

For Bernal stacked tBLG device, $S_{\mr{Mott}}$ is evaluated from numerically differentiating the measured $G_\mr{cp}$ with respect to gate potential $V_\mr{tg}$ and using the parabolic dispersion of the bilayer graphene \cite{mccann2006asymmetry,nam2010thermoelectric},

\begin{equation}
E(k)=\frac{1}{2}\gamma_1 \Big[\sqrt{1+\frac{v_{\mr{F}}^2\hbar^2 k^2}{\gamma_1^2}} -1 \Big].
\label{BLG dispersion}
\end{equation}
where $\gamma_1 \approx 0.39$~eV is the inter-layer hopping energy and $v_\mr{F} \approx 0.95 \times 10^6$ is the Fermi velocity in BLG. The BLG dispersion yields, $\frac{dV_{\mr{tg}}}{dE} = \frac{4}{\xi \gamma_1 } \sqrt{1 + \xi|V_\mr{tg}-V_\mr{D}|} $ where the factor $\xi = \frac{4v_\mr{F}^2 \hbar^2 \pi C_\mr{BN}}{e\gamma_1^2} \approx 1$ for gate-dielectric (hBN) of thickness $\approx 7$~nm. An excellent quantitative agreement between measured TEP and Mott relation was obtained (Fig.~4d) by scaling the $S_{\mr{Mott}}$ to compensate for the overestimation of the measured $\Delta T$ from resistive thermometry (see supplementary information). We verify that the similarly normalized $S_\mr{Mott}$, when evaluated from the single layer Dirac dispersion, does not conform well with the density variation of measured $S$ (inset of Fig.~4d). This validates that the band dispersion is non-Dirac and parabolic in the hybridised overlap region which suggests that the doping dependence of $S$ is highly sensitive to the coherence of the electronic states via the band dispersion (see supplementary information). Furthermore, we compare the electronic component of the TEP, $S_{\mr{Mott}}$ to the measured $S$ for large twist angle, $\theta \sim 12.5^{\circ}$ in the inset of Fig.~3e at low temperature ($30$~K) . The observation of large discrepancy from Mott relation even at low temperature $\sim 30$~K where $T/\mi{\Omega}_{\mr{ph}} \ll 1$, strongly suggests that the purely electronic part of the TEP is absent at large twist angle. 

\begin{figure}[t]
\includegraphics[clip,width=9cm]{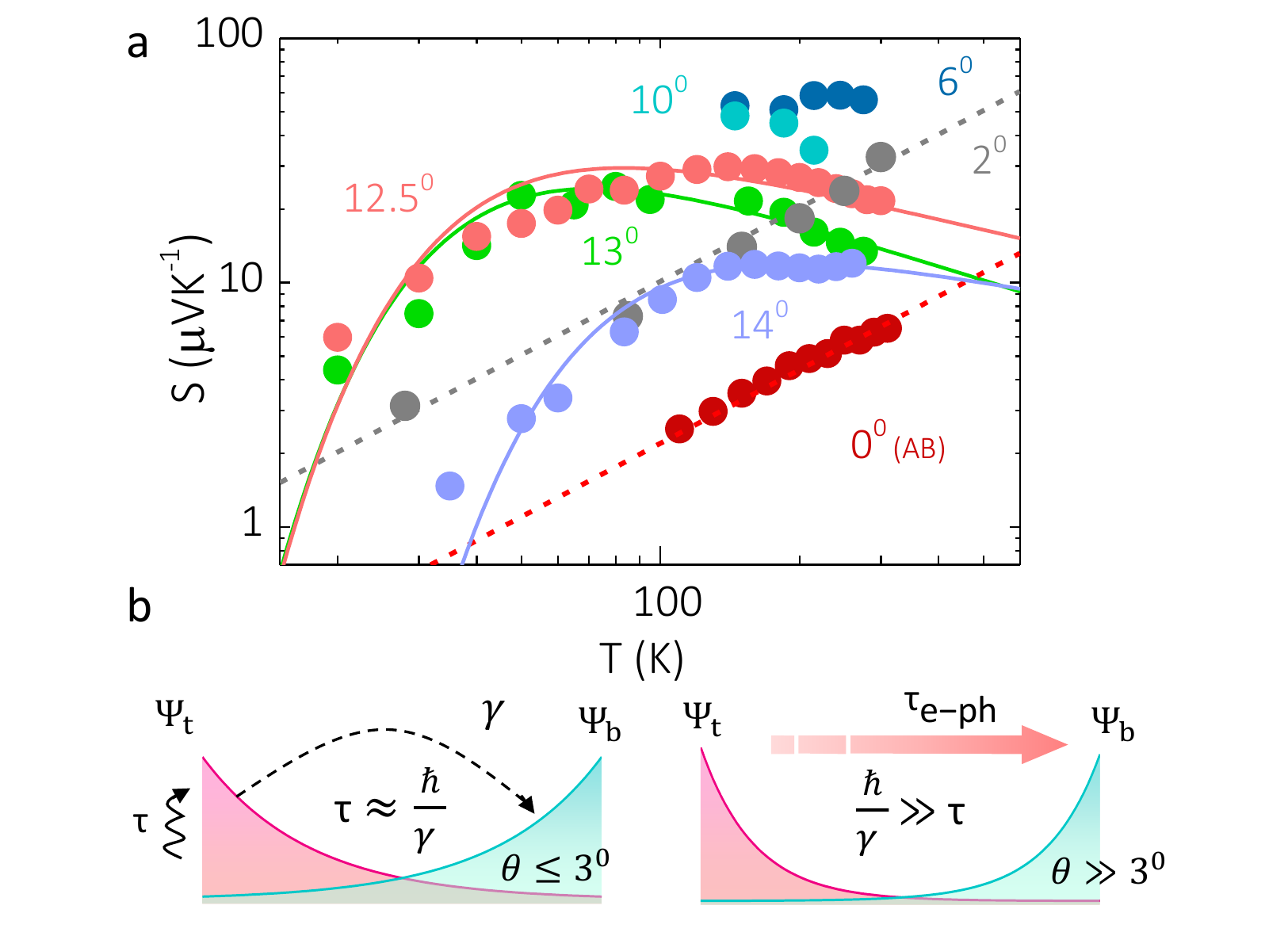}\caption{\textbf{Twist controlled cross-plane thermoelectricity}: (a) Seebeck coefficient as a function of temperature for seven different $\theta$ at a fixed doping of $n \sim 8 \times 10^{11}$~cm$^{-2}$. The solid lines show the fit of the phonon-driven TEP while the dotted lines show the linear $T$-dependence as a guide for the eye. The $S-T$ data for $\theta \sim 13^{\circ}$ is taken from Ref.~\cite{mahapatra2017seebeck}. (b) A schematic showing the interplay of the time scales associated with inter-layer hybridization, de-phasing and electron-phonon scattering. }
\end{figure}

In Fig.~5a, we present the $T$-dependence of all devices at a fixed representative number density $n \sim 8 \times 10^{11}$~cm$^{-2}$ which is close to the CNP. We observe that the large twist angle devices ($\theta \sim 6^{\circ}- 14^{\circ})$ show qualitatively similar non-monotonic TEP which is identified with the phonon drag mediated TEP. In contrast to large $\theta$, the devices with low twist angle ($\theta = 0^{\circ}$ and $\theta \sim 2^{\circ}$ ) exhibit linear $T$-dependence of $S$, indicating a crossover from phonon-driven thermoelectric transport to the purely coherent electronic transport. This striking shift of TEP from electronic hybridization to phononic hybridization with increasing twist angle demands further elaboration. We begin with estimating the generic e-ph scattering time scale $\tau_{\mr{e-ph}}$ from the phonon energy $\mi{\Omega}_{\mr{ph}}$ obtained from the intercept in Fig.~3d inset. As the inter-layer e-ph scattering requires a phonon to be absorbed/emitted, the time scale $\tau_{\mr{e-ph}} \sim \hbar/\mi{\Omega}_{\mr{ph}}$ can be estimated to be $\sim 10$~ps for $\mi{\Omega}_{\mr{ph}} \sim 100$~K. When the twist angle is reduced, $\tau_{\mr{e-ph}}$ shows weak dependence on $\theta$ due to the quadratic dispersion of $\mi{\Omega}_{\mr{ph}}$ . However, when the twist angle is increased, the interlayer electronic tunnelling time scale $\hbar/\gamma$ decays rapidly and becomes slower than $\tau_{\mr{e-ph}}$ for twist angle $\theta >6^{\circ}$ (Fig.~5b) \cite{bistritzer2010transport}. Consequently, e-ph scattering becomes the dominant mode of inter-layer charge transport instead of the electronic tunnelling at large $\theta$. The observed phonon-drag TEP even at $T \sim 30$~K$\ll \mi{\Omega}_{\mr{ph}}$ when e-ph scattering is not expected to be dominant, is seemingly due to the e-ph scattering length becoming comparable to mean free path or the cross-plane distance between the two layers which is not unfamiliar in clean graphene samples \cite{Kimhydro}. However, when mismatch is reduced to $\theta \leq 2^{\circ}$, the strong inter-layer hybridization drives the system to coherent tunnelling regime where the cross-plane tunnelling time scale ($\hbar/\gamma \sim 10-100$~fs) is expected to be much faster \cite{bistritzer2010transport} than the $\tau_{\mr{e-ph}} (\sim 10$~ps), effectively dominating any phonon contribution in the cross-plane transport and leading to $S \sim T/T_\mr{F}$.

In summary, we have experimentally measured the cross-plane thermoelectricity across a single van der Waals gap between two rotated graphene layers with varying twist angle. The measured Seebeck coefficient exhibits unique dependence on the twist angle and hence on the hybridization of electronic and phononic bands of two graphene layers. At large twist angle, the cross-plane thermoelectric transport is entirely driven by the e-ph scattering from the hybridized phonons which give rise to an unconventional phonon drag effect at the subnanometer distance, while at low twist angle ($\theta \leq 2^{\circ}$), the electronic hybridization is restored, resulting in a thermopower that can be described by the semiclassical Mott relation for coherent charge tunnelling. The twist-controlled thermoelectricity can not only probe the inter-layer coherent states in twisted bilayer graphene but may trigger new thermoelectric designs.

We thank HR Krishnamurthy, Manish Jain, Tanmoy Das and Sumilan Banerjee for valuable discussions. The authors thank DST for financial support and CeNSE for providing fabrication and characterization facilities. K.W. and T.T. acknowledge support from the Elemental Strategy Initiative conducted by the MEXT, Japan and the CREST (JPMJCR15F3), JST.

P.S.M. and B.G. contributed equally to this work.

$^{\ast}$ phanis@iisc.ac.in
$^{\#}$ gbhaskar@iisc.ac.in

\bibliographystyle{naturemag}
\bibliography{TBLG}

\end{document}